\documentclass[11pt,showpacs,nofootinbib]{revtex4-1}
\usepackage{epsfig}
\usepackage{amssymb}
\usepackage{amsmath}
\usepackage{amsthm}
\usepackage{amsopn}

\def\comment#1{}

\def\beq{\begin{equation}}
\def\eeq{\end{equation}}
\def\bea{\begin{eqnarray}}
\def\eea{\end{eqnarray}}

\usepackage{ulem}
\usepackage{cancel}
\usepackage{color}

\begin{document}
    {\textsf{\today}
        \vspace*{1cm}
\title{Are latest detected events of gravitational waves in favor of some models of inflation based on string theory?}
\author{ Basem Ghayour$^{1}$\footnote{ba.ghayour@gmail.com}, Jafar Khodagholizadeh$^{2}$\footnote{j.gholizadeh@cfu.ac.ir},
M. Afkani$^{3}$\footnote{std\_ m.afkani@khu.ac.ir} and M. R. Torkamani$^{3}$\footnote{std\_ mr.torkamani@khu.ac.ir}, Ali Vahedi$^{3}$\footnote{vahedi@khu.ac.ir}}  
\affiliation{$^{1}$ School of Physics, University of Hyderabad, Hyderabad 500 046, India.\\    
    $^{2}$ Farhangian University, P.O. Box 11876-13311,  Tehran, Iran.\\
    $^{3}$ Department of Physics, Kharazmi University, P.O. Box 15614, Tehran, Iran.}

%\maketitle

\begin{abstract}
The general potential of power-law inflation is as $ V(\phi)\propto \phi^{n} $ with scalar field $\phi$. The behavior of inflation is often known as power-law expansion like $S(\eta)\propto \eta^{1+\beta}$ with $1+\beta<0$. In this paper, the theoretical spectra of relic gravitational waves are compared with the measured strain sensitivity of Advanced LIGO and VIRGO, corresponding to the latest detected events of gravitational waves. The results show tight constraints on $\beta$ and $n$. Also, the obtained constraints indicate that special types of the potential of inflation, prototype, and KKLTI models, which are originated from string theory, may be good candidates for potential of inflation. Also these results may emphasize the evidence of stochastic GWs that originated from inflation based on 12:5-year pulsar timing data set of North American Nanohertz Observatory for GWs.

\end{abstract}

\pacs{13.15.+g, 98.80.Es, 98.70.Vc.}

\maketitle
%\newpage
\section{Introduction}    
Gravitational waves are a generic prediction of inflation in the early universe \cite{Guth, Linde, Albrecht}. In comparison, the relic gravitational waves (RGWs) and stochastic GWs are generated during  inflation stage \cite{Grishchuk, Grishchuk1, Grishchuk2, Grishchuk3, Starobinsky, Starobinsky1, Rubakov}. Therefore, it seems that inflation is the main source of GWs. To date, there has been no exact form of producer potential of inflation. Recently, Advanced LIGO (Adv.ligo) \cite{ALIGO, ALIGO1} and Virgo \cite{VIRGO, VIRGO1} detectors have been listed  the latest detected events of GWs. These events are called GW150914 to GW170823 (GWGs) \cite{el}. Also there is a search for an isotropic stochastic gravitational-wave background  in the 12:5-year pulsar
timing data set collected by the North American Nanohertz Observatory for GWs
(NANOGrav) \cite{qwq}. This analysis finds strong evidence of a stochastic GWs, modeled as a power-law,
with common amplitude and spectral slope across pulsars. As one of the main source of stochastic GWs in the band $\sim (1-100)$ nHz is inflation \cite{qwq}. In our previous work \cite{Ghayour}, we showed that there exist some chances for detecting the theoretical spectrum of RGWs including thermal spectrum in addition to the usual spectrum by comparing the strain sensitivity of Adv.LIGO for GW150914. Similarly, there is corresponding measured strain sensitivity in the range $ \sim (10^{-1}-10^{4}~)$ Hz of the Adv.ligo and Virgo detectors during the time analyzed to determine the significance of GWGs \cite{el}. The comparison of the theoretical spectrum with the measured strain sensitivity of GWGs can  give us some results. These results provide valuable information about the form of the potential of inflation and the evolution of the waves. For more details in this regard, please see \cite{Ghayour}. Also these results may emphasize the evidence of stochastic GWs that originated from inflation based on work \cite{qwq}. Therefore, it would be interesting to study thsee results. 

There are different classes of inflationary models:
(1) Large field such as polynomial, power-law inflation \cite{Liddle6,Bauman, Langios}, and chaotic inflation models \cite{Lewis} and ~(2) a small field such as Hiltop inflationary \cite{Boubekeur} and chaotic models \cite{Martin4}.
Also, there is another model that is based on cosmological perturbation theory in the Brane-World gravity, which is widely ruled out by Planck \cite{Liddle6, Langios, Martin3, Martin4, Lewis}. Based on string theory, researchers have proposed different models such as the prototype model \cite{Port}, KKTLI model \cite{KK}, and IR Dirac-Born-Infeld(DBI) model \cite{DBI}. The prototype model of brane inflation and KKTLI are still in good agreement with the direct $H_{0}$ measurement \cite{Rui} and Planck 2018 \cite {Planck2018}, respectively. 
\\
The general potential of power-law inflation is like $ V(\phi)\propto \phi^{n} $ with scalar field $\phi$. Also, the behavior of inflation is often known as power-law expansion like $S(\eta)\propto \eta^{1+\beta}$ with constraint $1+\beta<0$, where $S$ and $\eta$ are scale factor and conformal time, respectively. For a given value of the spectral index $ n_{s} $, it has been shown that the general range on $n$ based on Planck 2018 data  \cite {Planck2018} is ($n<1$) and then its corresponding  range on $\beta$ can be found. 
%(see ~\cite{to} for more details).

However, comparing the theoretical spectrum of RGWs and measured strain sensitivity of GWGs indicates that there  are some interesting ranges on $n$ and $\beta$. Our obtained range on $n$ in this work is $(-3.5 \lesssim n \lesssim -1.8)$ and on the corresponding range of $\beta$ would be $(-2.035\lesssim \beta \lesssim-1.814)$. As the obtained most sensitive upper limits of frequency band
$\sim (20-100)$ Hz in recent paper \cite{pm} is within the frequency band $\sim (10^{-1}-10^{4})$ Hz of the correspond obtained ranges of $\beta$ in our work.  There is  an acceptable range $ -2 \lesssim \beta \lesssim  -1.814$ based on
works in \cite{Grishchuk2, Ghayour}. As it seems  that our obtained range on $n$ ($n<0$) is consistent with general range $n<1$ and also the corresponding range on $\beta$ covers the acceptable range.
%But from other hand, the range with $n>0$ is also compatible with Adv.ligo and Virgo data \cite{Planck2018}. 
Therefore it expects that there is an ambiguity and  it  depends on which physics  that one is going to use for describing $n>0$ or $n<0$. Based on the measured strain sensitivity, we may conclude that the case with $n< 0$ is more suitable than the case with $n>0$ corresponding to GWGs.  Therefore, we are interested to challenge the range with   $n>0$  in favor of $n<0$ in this work. This new constraint is in agreement with the prototype and KKTLI inflationary models, which are still consistent with data \cite{Port, KK, Rui, Planck2018}. Hence, our results suggest that the special string theory models of inflation (prototype and KKLTI) may be a good candidates for  potential of inflation. In the present work, we use the unit $c = \hbar = k_{B} = 1$.

The remainder of this paper is organized as follows: In section II, there is a brief review of RGWs in the universe. In Section III, constrained parameters $n$ and $\beta$ is obtained based on GWGs. In Section IV, we discuss the corresponding inflation in Brane-World gravity in our results. Finally, in Section 5, the conclusion is presented along with relevant discussions and comparisons of the results.

\section{RGWs in the expansion universe}
 In the Friedmann-Robertson-Walker universe, $ h_{ij} $ are tensor mode of metric perturbations with the transverse-traceless properties, i.e., $ \triangledown_{i}h^{ij}=0=h^{i}_{~i} $. The linearized gravitational waves equation is 
\begin{equation}\label{equation}
\triangledown_{\mu}(\sqrt{-g}\triangledown^{\mu}h_{ij}(\boldsymbol{x},\eta))=0
\end{equation}
where $ \eta $ is the conformal time. The tensor mode perturbations have two-mode polarization that can be expressed in terms of the creation $ a^{+} $ and annihilation $ a $ operators
\begin{equation}
h_{ij}(\boldsymbol{x}, \eta)=\dfrac{\sqrt{16\pi}l_{pl}}{a(\eta)}\sum_{\sigma}\int \dfrac{d^{3}k}{2 \pi^{3/2}} \epsilon_{ij}^{\sigma}(\boldsymbol{k})\dfrac{1}{\sqrt{\sqrt{2k}}}[a_{\boldsymbol{k}}^{\sigma}h_{\boldsymbol{k}}^{\sigma}(\eta)e^{i\boldsymbol{k}.\boldsymbol{x} }+a_{\boldsymbol{k}}^{^{\dagger}\sigma}h_{\boldsymbol{k}}^{^{\ast}\sigma}(\eta)e^{-i\boldsymbol{k}.\boldsymbol{x} }]
\end{equation}
where $ \boldsymbol{k} $ is the comoving wave number with $ k=|\boldsymbol{k}| $, $ l_{pl} $ is the Planck length, and $ \sigma=+ $ and $ \times $ are polarization modes. The $ \epsilon_{ij}^{\sigma}(\boldsymbol{k}) $ is the polarization tensors with the symmetric condition, $ \delta^{ij}\epsilon_{ij}^{\sigma}(\boldsymbol{k})=0 $ and transverse-traceless property $ k^{i}\epsilon_{ij}^{\sigma}(\boldsymbol{k})=0$ satisfy these conditions \cite{Grishchuk1}.
\begin{eqnarray}
{\epsilon^{ij~\sigma^{}}(\boldsymbol{k})\epsilon_{ij}^{\sigma^{'}}(\boldsymbol{k})}&=&2\delta_{\sigma\sigma^{'}},\\\epsilon_{ij}^{\sigma}(\boldsymbol{-k})&=&\epsilon_{ij}^{\sigma}(\boldsymbol{k})\,.
\end{eqnarray}
For a fixed wave number $ \boldsymbol{k} $ and a fixed polarization state $ \sigma $, Eq.(\ref{equation}) gives
\begin{equation}\label{eq:1}
h_{k}^{(\sigma)''}(\eta)+2\dfrac{a^{'}(\eta)}{a(\eta)}h_{k}^{(\sigma) '}(\eta)+k^{2}h_{k}^{(\sigma)}(\eta)=0\,\footnote{where a prime means taking derivative with respect to $\eta$ .} ,
\end{equation}
where the analytical solutions of this equation could be found in \cite{Miao}. The scale factor of the inflation stage would be $ a(\eta)\propto \mid\eta\mid^{1+\beta} $, which means that the $\beta$ plays the main role on the shape of the spectrum of RGWs \cite{Grishchuk2}. A constraint has been obtained on the $\beta$ from theoretical model $\beta<-1$ \cite{Grishchuk2, Tong} and also by corresponding observation $\beta\lesssim-1.804$ \cite{Ghayour}. 
\section{ constrained parameters  based on GWGs  }
The general potential of power-law inflation is like $V(\phi) \propto \phi^{n}$ with scalar ﬁeld $ \phi $. The results are consistent with inflation equations with the Hubble parameter $ H $ as follows
\begin{equation}
\ddot{\phi}+3 H \dot{\phi}+\dfrac{d V}{d\phi}=0~~~~~,~~~~~ H^{2}=\dfrac{1}{3}[\dfrac{1}{2}\dot{\phi}^{2}+V(\phi)]
\end{equation}
This parameter can easily be solved by slow-roll approximation. Under the slow-roll conditions, the evolution of inflation is described by two parameters \cite{Liddle, Liddle1, Liddle2}
\begin{equation} \label{epsilon}
\epsilon=\dfrac{m_{Pl}^{2}}{16\pi} (\dfrac{V^{'}}{V})^{2} ~~~~~,~~~~~\eta= \dfrac{m_{Pl}^{2}}{8\pi} \dfrac{V^{''}}{V}
\end{equation}
where the prime stands for derivative of potential with respect to $\phi$. These quantities are smaller than unity. They are dimensionless and especially $ \epsilon $ approaches to unity at the end of inflation. Also, the primordial tensor power spectrum and the scalar power spectrum are given as \cite{, Kuro, Boyle}
\begin{eqnarray}\label{ratio}
\Delta_{h}^{2}(k, \eta_{\star})&\approx& \dfrac{16}{\pi}(\dfrac{H_{\star}}{m_{Pl}})^{2}\nonumber \\
\Delta_{R}^{2}(k, \eta_{\star})&\approx& \dfrac{1}{\pi \epsilon}(\dfrac{H_{\star}}{m_{Pl}})^{2}
\end{eqnarray}
respectively, where $ H_{\star} $ is the Hubble rate during inflation and $ \eta_{\star} $ stands for the moment when the k mode exits the horizon. In addition, based on the observations of CMB, the present tensor and scalar power spectrum can be expanded in power laws
\begin{eqnarray}
\Delta_{h}^{2}(k)&=&\Delta_{h}^{2}(k_{0}) (\dfrac{k}{k_{0}})^{n_{t}}\nonumber \\
\Delta_{R}^{2}(k)&=& \Delta_{R}^{2}(k_{0}) (\dfrac{k}{k_{0}})^{n_{s}-1}
\end{eqnarray}
where $ \Delta_{h}^{2}(k_{0}) $ and $ \Delta_{R}^{2}(k_{0}) $ are evaluated at the pivot wave number $ k_{0}^{p}= k_{0}/a(\eta_{0}) = 0.002 Mpc^{-1}$ \cite{Komatsu}, respectively. There is a relation between tensor index $n_t$ and $\beta$ as follows
\begin{equation}\label{fffd}
n_t=2\beta +4,
\end{equation}
and also between
 $\beta$ and $n$ based on $V\propto \phi^{n}$ as ~\cite{to} (see Appendix A for more details):   
\begin{figure}
   \includegraphics[scale=0.24]{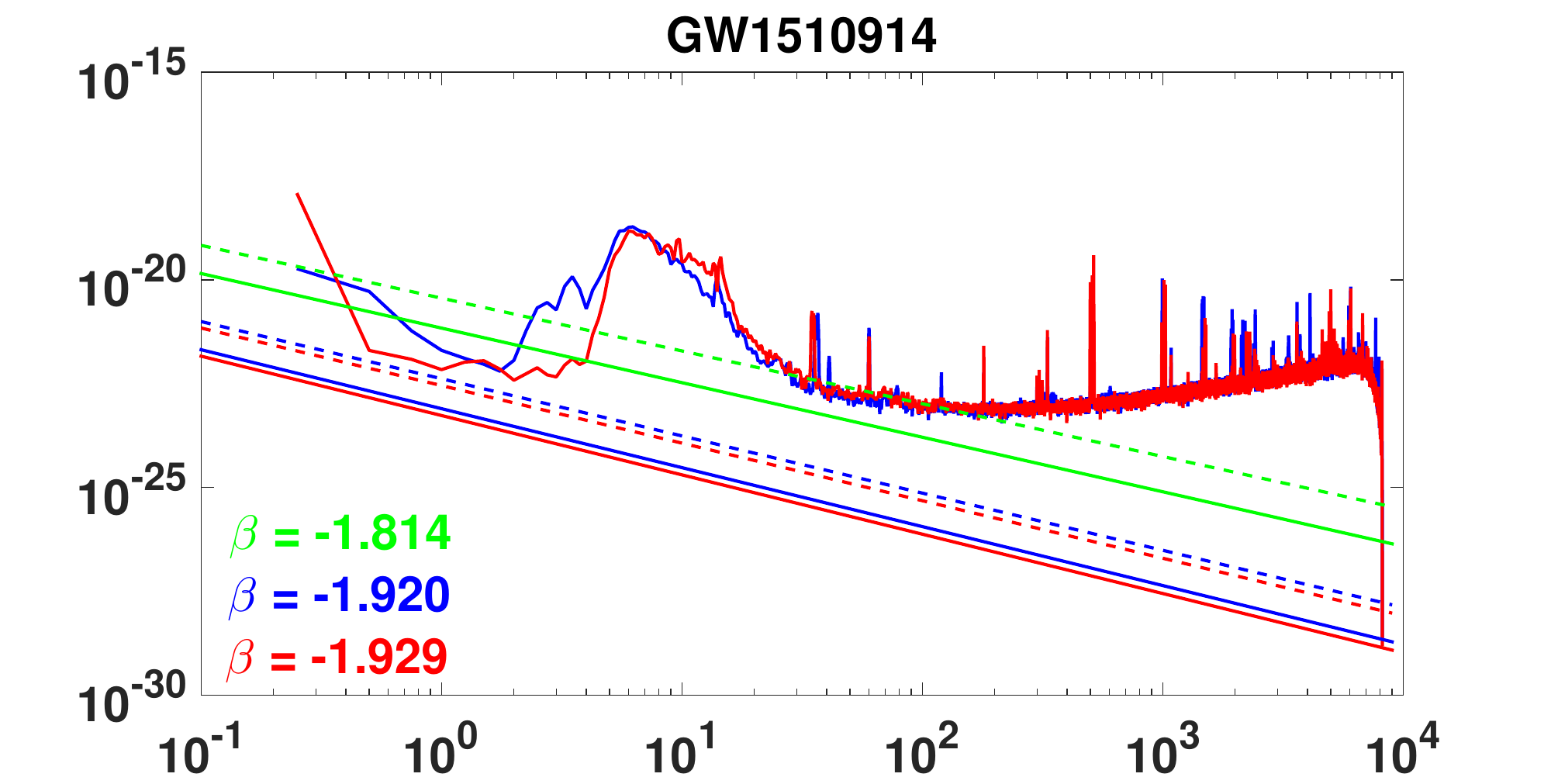}
   \includegraphics[scale=0.24]{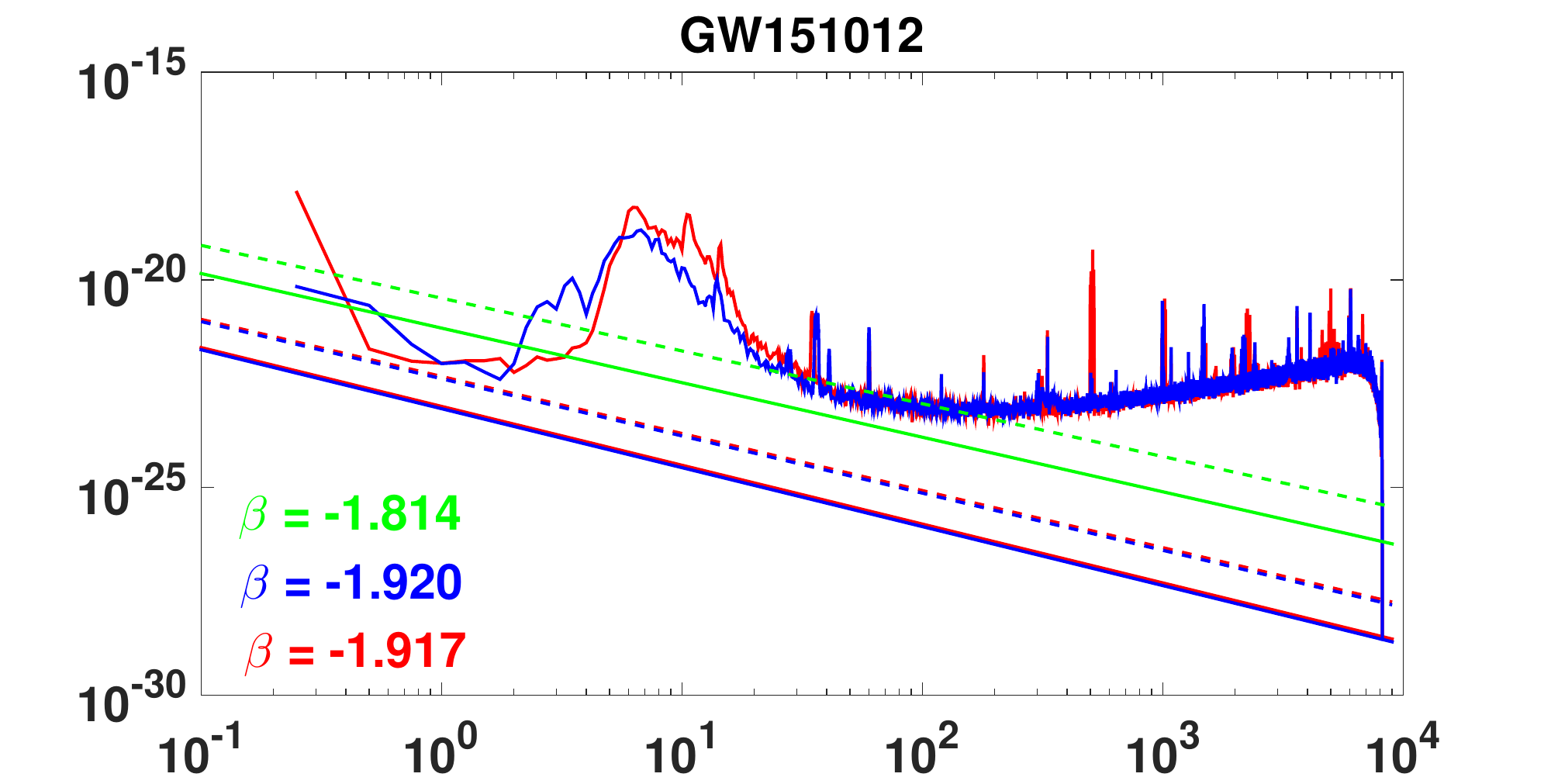}
      \includegraphics[scale=0.24]{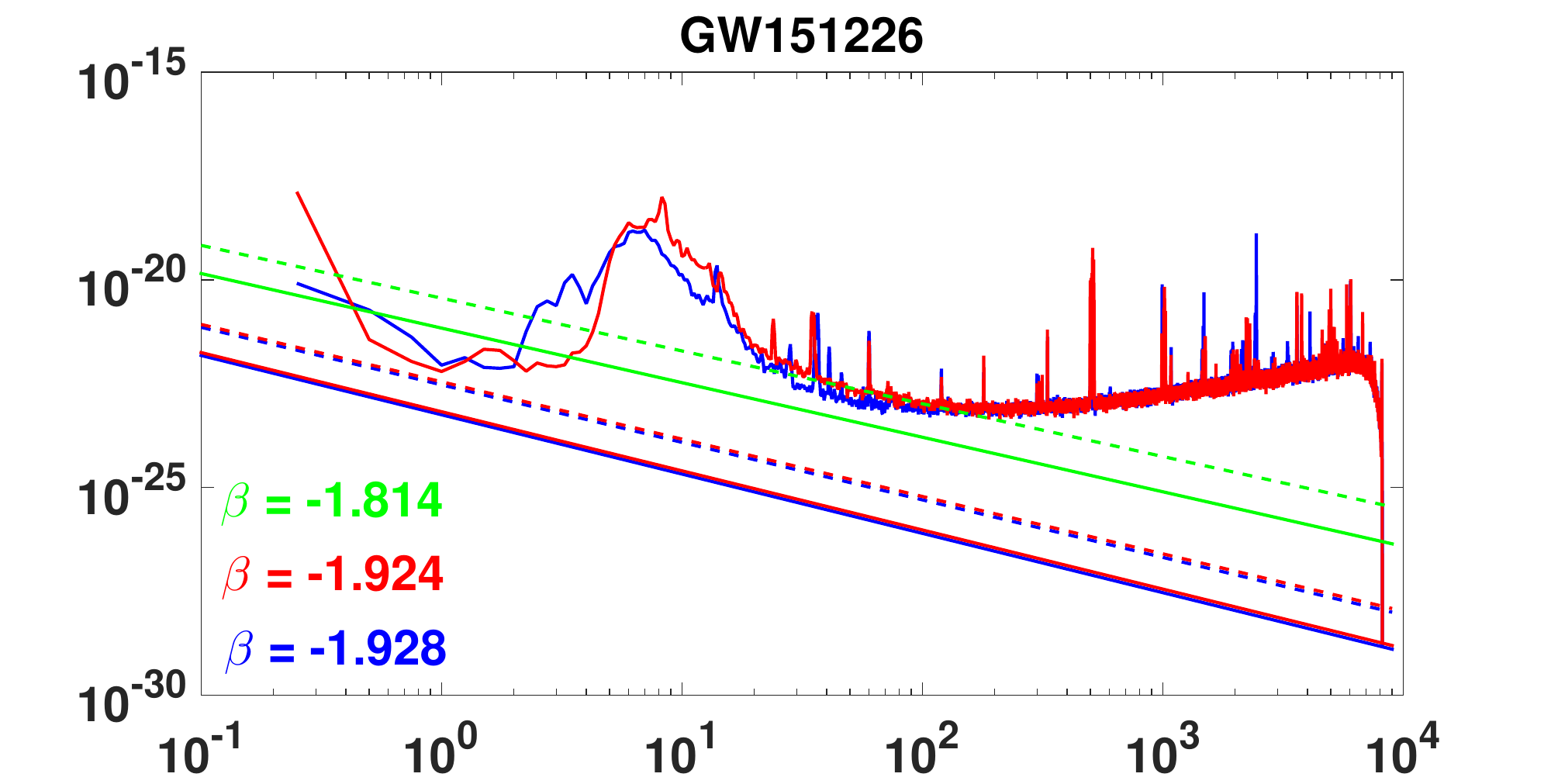}
 \includegraphics[scale=0.24]{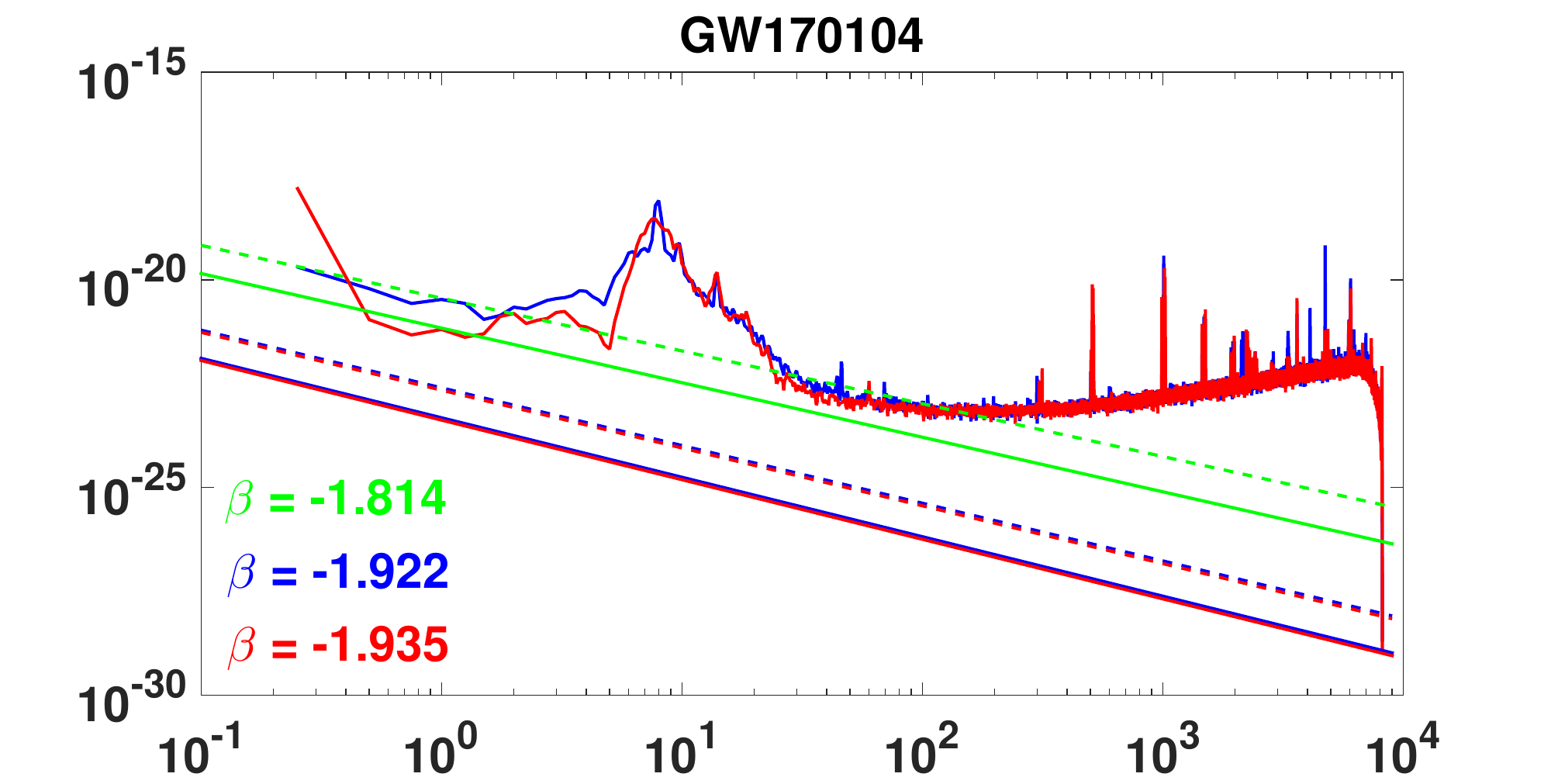}
   \includegraphics[scale=0.24]{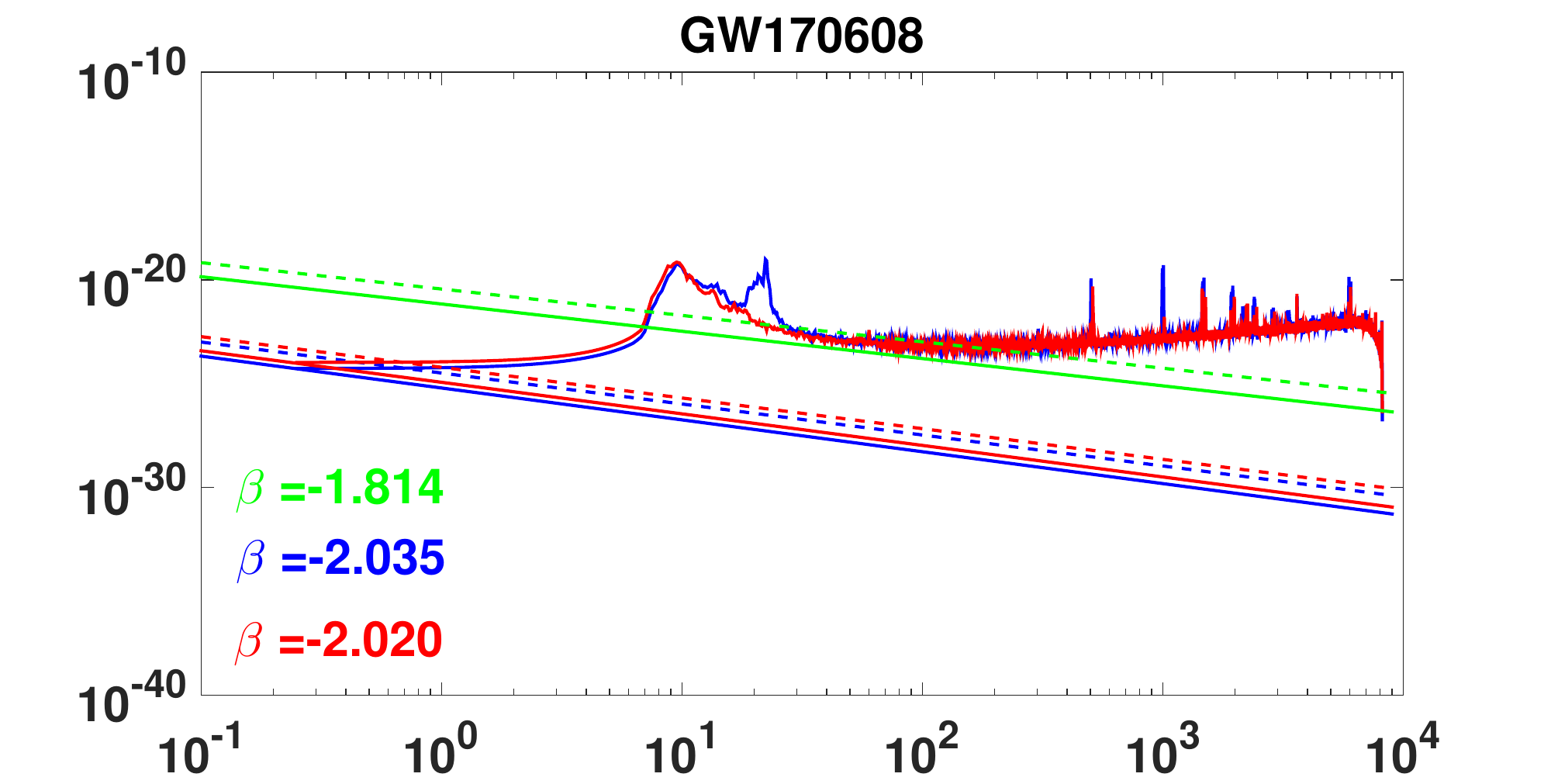}
      \includegraphics[scale=0.24]{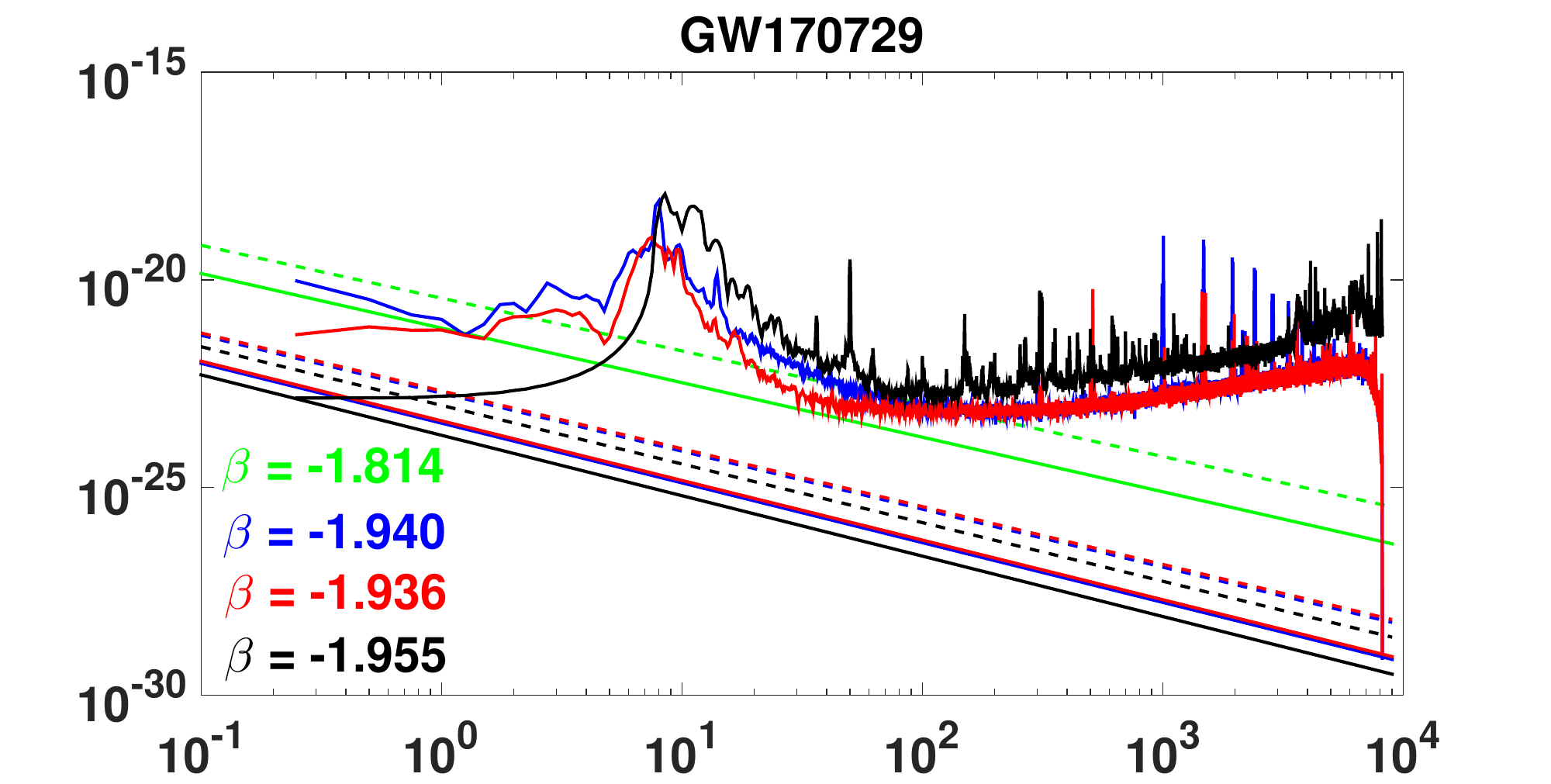}
   \caption{ The theoretical spectrum of RGWs that contains thermal spectrum \cite{Ghayour} (dashed lines) and usual
   	spectrum (solid lines) compared to corresponding measured strain sensitivity of Adv.ligo (Hanford, blue color and Livingston, red color) and Virgo (black color) during the time analyzed to determine the signiﬁcance of GWGs \cite{el}. 
   	The green line shows the upper bound on $\beta\lesssim-1.814$ \cite{Ghayour}. Also, the blue, red, and black lines show the lower bound on $\beta$ compared to Hanford, Livingston, and Virgo, respectively. Notably, in each panel, the horizon and vertical axes stand for frequency (Hz) and strain sensitivity (per root Hz), respectively. 
    }
    \label{fig}
\end{figure}

  \begin{figure}
    \includegraphics[scale=0.24]{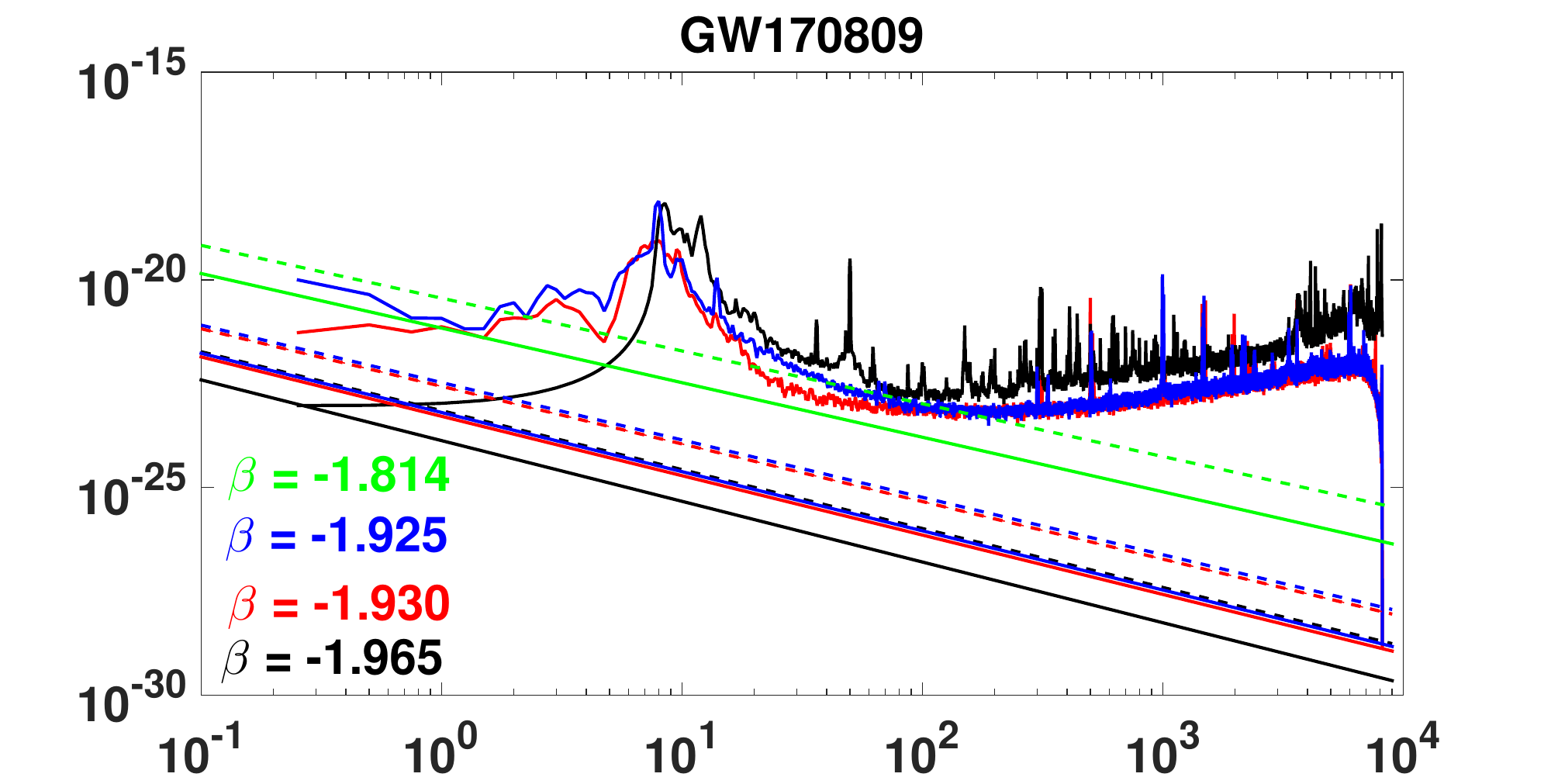}  
     \includegraphics[scale=0.24]{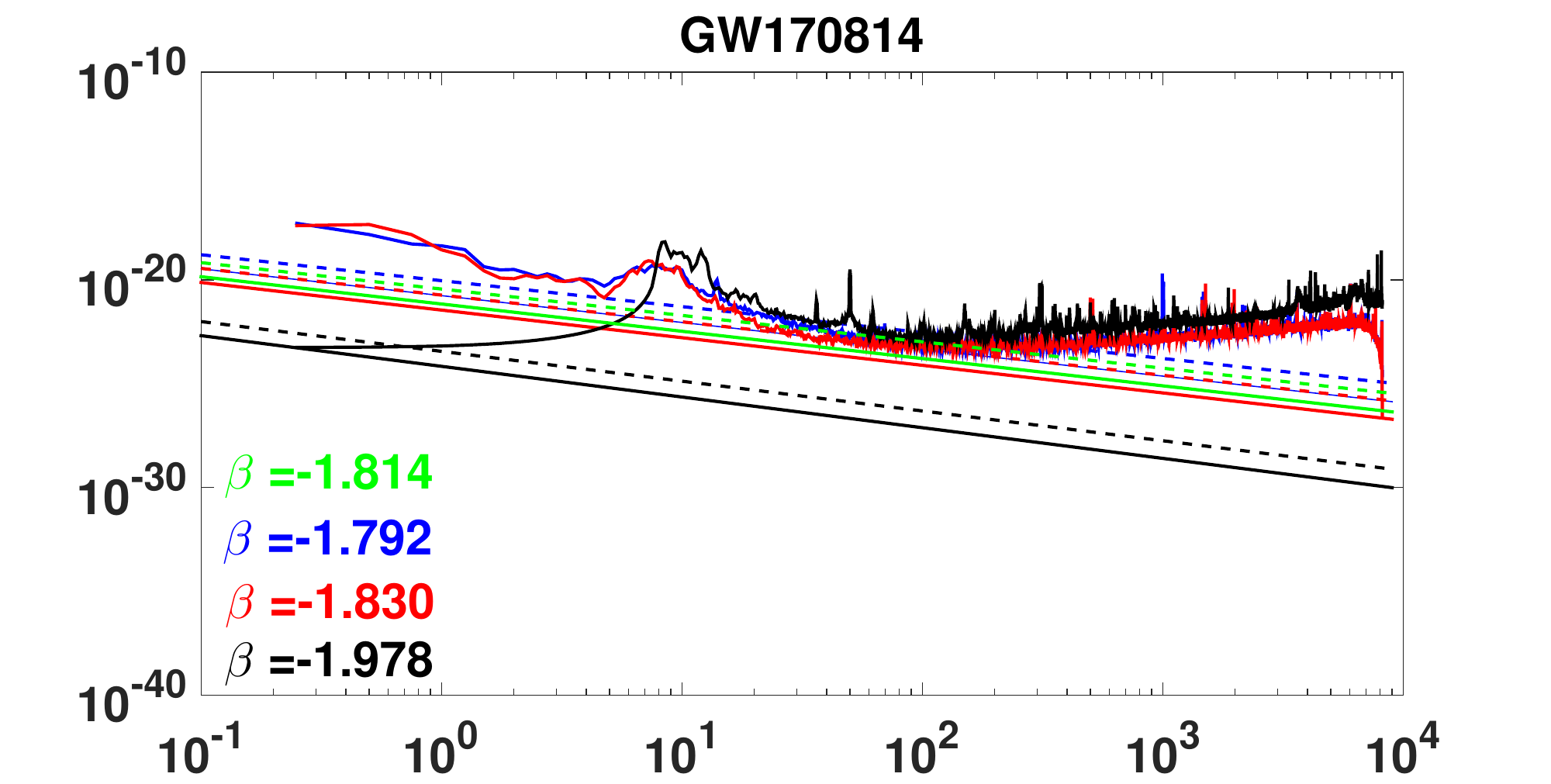} 
      \includegraphics[scale=0.24]{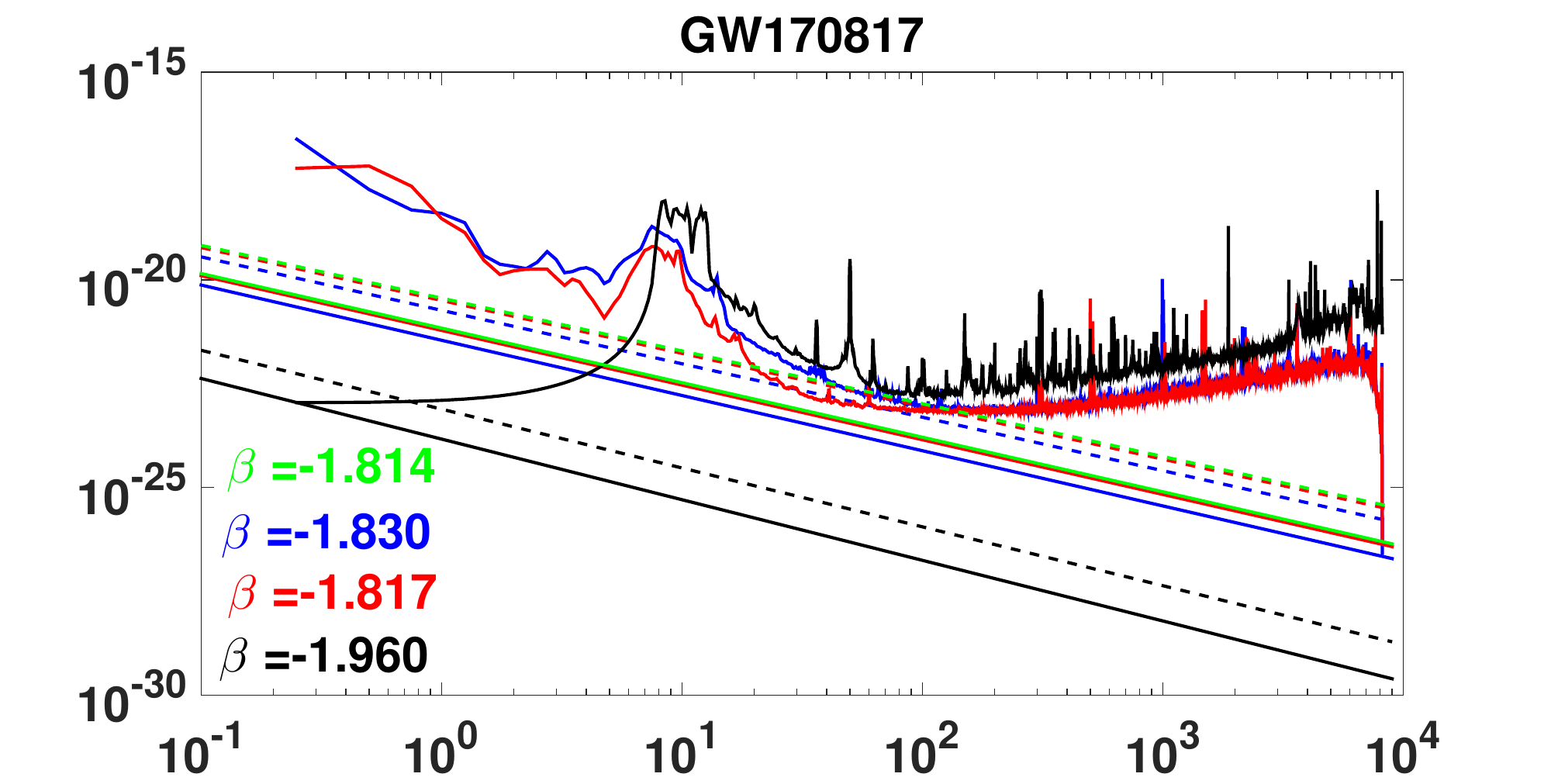}
         \includegraphics[scale=0.24]{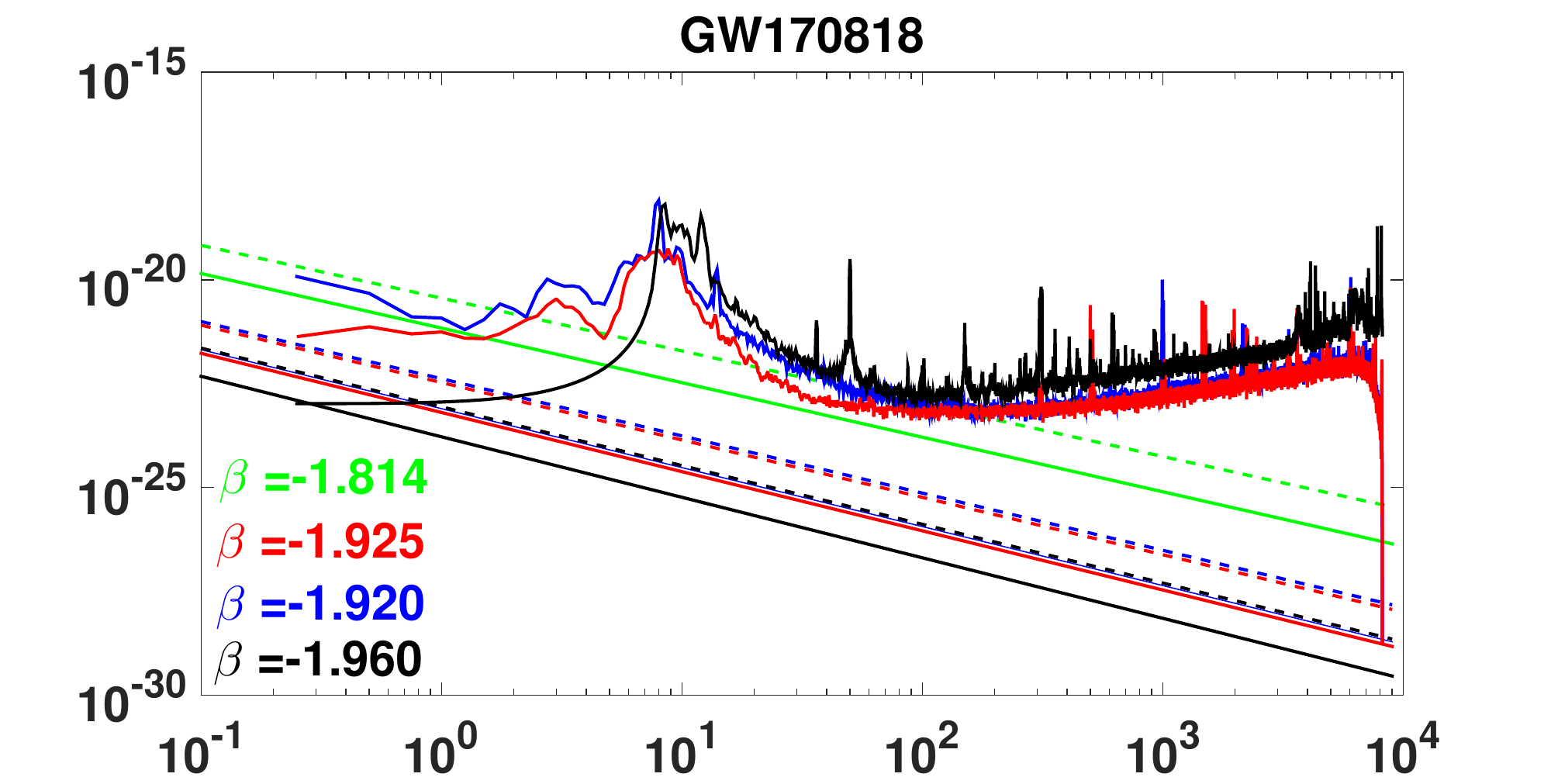}
            \includegraphics[scale=0.24]{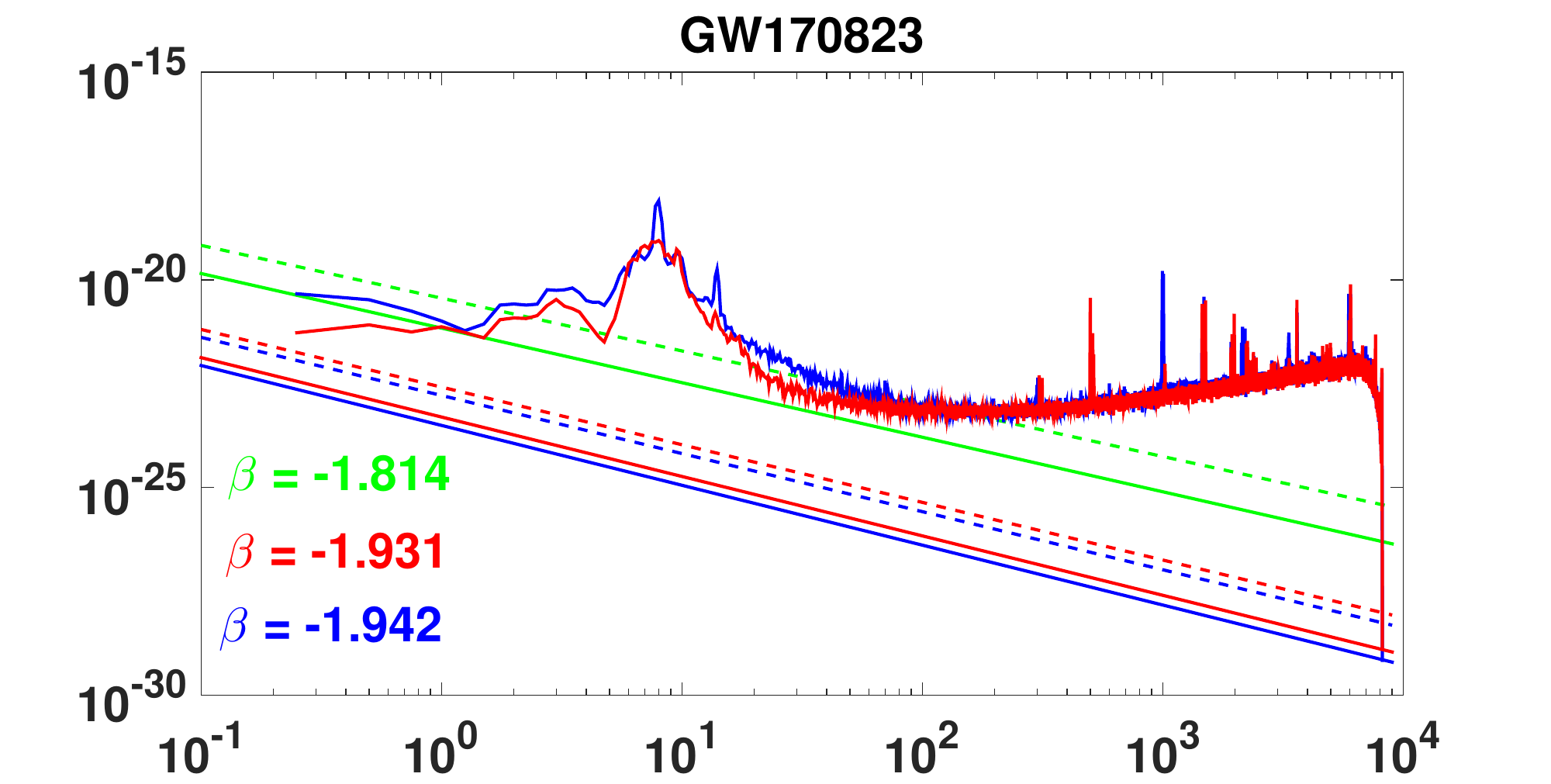}
    \caption{Continued fig.1.   }
    \label{figg}
\end{figure}
\begin{equation}\label{betaa}
\beta=-2-\dfrac{n}{2(n+2)}(1-n_{s})
\end{equation}
where the parameter $ n $ is constrained based on Planck 2018 ~\cite{Planck2018} in the range 
\begin{equation}\label{dc}
n<1,
\end{equation}
and then corresponds range on 
 $ \beta $ will be found for given $ n_{s} $.
  For example with $n_{s}=0.97$, it will be
 	\begin{equation}\label{cds}
 -2.005<\beta.
 	\end{equation}  
 %In fact, these constraints are consistent with theory of inflation and CMB \cite{to}. 

\begin{table}
    \begin{center}
        \begin{tabular}{ | c | c  | c | c |c | }\hline
           GWGs  ~~&      ~~Hanford ~~ & ~~Livingstone  ~~&      ~~{Virgo}  ~~&      ~~range ~~\\ \hline
            GW150914 & $  -1.920\lesssim\beta   $&  $-1.929\lesssim \beta  $ & $---- $ & $ -1.929\lesssim\beta \lesssim-1.814 $ \\  \hline
            GW151012 &   $ -1.920\lesssim\beta  $& $  -1.917\lesssim\beta   $&  $ ---- $ & $ -1.920 \lesssim\beta \lesssim-1.814  $  \\ \hline
            GW151226 &    $ -1.928\lesssim\beta $& $  -1.924\lesssim\beta   $&  $---- $ & $-1.928 \lesssim\beta \lesssim-1.814 $   \\ \hline
            GW170104 &  $ -1.922\lesssim\beta  $& $ -1.935\lesssim\beta  $&  $ ----  $ & $ -1.935\lesssim\beta \lesssim-1.814  $    \\ \hline
            GW170608 &  $-2.035\lesssim\beta  $& $  -2.020\lesssim\beta   $&  $----  $ & $  -2.035\lesssim\beta \lesssim-1.814  $    \\ \hline
            GW170729 &   $ -1.940\lesssim\beta$& $  -1.935\lesssim\beta   $&  $ -1.955\lesssim\beta  $ & $ -1.955 \lesssim\beta \lesssim-1.814  $   \\ \hline
            GW170809 &    $ -1.925\lesssim\beta $& $  -1.930\lesssim\beta   $&  $ -1.965\lesssim\beta   $ & $ -1.965  \lesssim\beta \lesssim-1.814  $  \\ \hline
            GW170814 &   $ -1.792\lesssim\beta $& $  -1.830\lesssim\beta  $&  $ -1.987\lesssim\beta  $ & $ -1.987 \lesssim\beta  \lesssim-1.814  $  \\ \hline
            GW170817 &    $ -1.830\lesssim\beta $& $ -1.817\lesssim \beta  $&  $-1.960\lesssim \beta  $ & $ -1.960 \lesssim\beta \lesssim-1.814  $     \\ \hline
            GW170818 &    $ -1.920\lesssim\beta $& $  -1.925\lesssim\beta  $&  $-1.960\lesssim \beta   $ & $ -1.960  \lesssim\beta\lesssim-1.814  $   \\ \hline
            GW170823 &    $ -1.942\lesssim\beta $& $  -1.931\lesssim\beta  $&  $----  $ & $ -1.942\lesssim\beta \lesssim-1.814 $   \\ \hline
        \end{tabular}
    \end{center}
    \caption{ The obtained bounds on $ \beta $  based on the measured strain sensitivity correspond to GWGs \cite{el}.
        Note that there is no released sensitivity of Virgo for some events. The different ranges on $\beta$ corresponds to each event lead to constraint    $ -2.035 \lesssim \beta \lesssim-1.814  $.\label{tab}}
\end{table}
In our previous study \cite{Ghayour}, we showed that there exist some chances for detecting the theoretical spectrum of RGWs that contains thermal spectrum in addition to the usual spectrum by comparison with strain sensitivity of Adv.LIGO for GW150914. Similarly, there is corresponding measured strain sensitivity in the range $ \sim (10^{1}-10^{4} )$ Hz of the Adv.ligo and Virgo detectors during the time that were analyzed to determine the signiﬁcance of GWGs \cite{el}.  %Also these events and RGWs emphasize that, there is a stochastic background throughout
%the history of the universe \cite{q1,q2,q3,q4}. 
The same as work in\cite{Ghayour}, in addition to more chance of detecting the RGWs by considering the comparison of the theoretical spectrum with the measured strain sensitivity, it seems that this comparison can give us other results as well.

The range in Eq. (\ref{dc}) and its corresponding range on $ \beta $ in  eq.(\ref{cds}) will modify due to measured strain sensitivity. In Fig. [\ref{fig}], we plot the theoretical spectrum of RGWs that contains thermal spectrum \cite{Ghayour2} (dashed lines) and usual spectrum (solid lines) compared to measured strain sensitivity of Adv.ligo (Hanford, blue color and Livingston, red color) and Virgo (black color) with the same used parameters in \cite{Ghayour}.
The green line shows the upper bound on $\beta\lesssim-1.814$ \cite{Ghayour}. Also, the blue, red and black lines show the lower bound on $\beta$ compared to Hanford, Livingston and Virgo respectively. Moreover, all obtained comparative values of $\beta$ are shown in Table (\ref{tab}). The last column of Table (\ref{tab}) shows the obtained range of $\beta$ for different events. Hence, these different ranges lead to a constraint on $\beta$ as follows:
  \begin{equation}\label{ee}
-2.035  \lesssim \beta \lesssim -1.814,
\end{equation}
  After ward, based on Eq. (\ref{betaa}) for  given $n_{s}\simeq0.97$ as a sample, we have
\begin{equation}\label{sa}
-3.5 \lesssim n \lesssim -1.8,
\end{equation}
It must be noted that the sign of $n$ does not change for  given $n_{s}\simeq0.96$ to $0.98$ [36]. Clearly the obtained range on $n_t$ from eqs.(\ref{fffd}, \ref{ee}) with
\begin{equation}\label{saaa}
-0.07  \lesssim n_t  \lesssim 0.37,
\end{equation}
 in this work is consistent with the range of Planck 2018 \cite{Planck2018} as
\begin{equation}\label{sadd}
-0.62  \lesssim n_t  \lesssim 0.53.
\end{equation}
Moreover that the obtained new ranges on $\beta, n$ in eqs.(\ref{ee},\ref{sa}) is in agreement with general range in  eqs.(\ref{dc},\ref{cds}) respectively.  Also   our obtained range in eq.(\ref{ee}) covers the acceptable range $-2 \lesssim \beta \lesssim -1.814$ based on works in \cite{Grishchuk2, Ghayour}. Hence it seems that the obtained range on $n$ is in favour of $n<0$ based on GWGs and can give some interesting result.  %But from other hand, the range with $n>0$ is also compatible with Adv.ligo and Virgo data \cite{Planck2018}. 
Therefore it expects that there is an ambiguity and  it  depends on which physics  that one is going to  use for describing $n>0$ or $n<0$ ! Based on the
measured strain sensitivity, we may conclude that the case with $n < 0$ is more suitable than the
case with $n > 0$ corresponding to GWGs. Thus we are interested to  challenge the range with   $n>0$ in eq.(\ref{dc}) in favor of $n<0$.  Therefore, we will try to justify this contradiction by introducing a suitable source of potential $\thicksim\phi^{n}$ for $n<0$ in the next section.

\section{Inflation in Brane-World Gravity}
Brane-world gravity is a theory of gravity in which space-time has $ (1+3+d) $ dimension with $ (1+3) $ brane embedded in $ (1+3+d) $ dimension bulk \cite{Marteens}. The standard model of particles (gauge theory) resides on the brane while gravity can live in the bulk. Here, we consider an inflationary dynamics in the brane-world in which the slow-roll parameters change as  
\begin{eqnarray}
\epsilon_{v}=\dfrac{M_{4}^{2}}{16\pi} (\dfrac{V^{'}}{V})^{2}\bigg[\dfrac{1+\dfrac{V}{\lambda}}{(1+\dfrac{V}{2\lambda})^{2}}\bigg]
\end{eqnarray}  
\begin{eqnarray}
\eta_{v}=\dfrac{M_{4}^{2}}{8\pi}(\dfrac{V^{''}}{V})\bigg[\dfrac{2\lambda}{2\lambda+V}\bigg]
\end{eqnarray}
where $ \lambda\geq (1 MeV)^{4} $ and $ M_{4} $ is the mathematical value of the Planck Mass in 4 dimensions as $ M_{4}=10^{27}eV $ (see \cite{Flanagan:1999cu} and references therein). The changes from standard General Relativity, based on calculations, is seen in the high energy as both the parameters are suppressed by a factor of $ \dfrac{V}{\lambda} $. \
In the brane-world cosmology, there are some models such as Large Field Inflation \cite{Linde1, Silverstein, Martin2}, Power Law Inflation \cite{Martin2, Abbot, Sahni, Sahni}, Open String Tachyonic Inflation \cite{Minahan}, and Inverse Monomiaal Inflation. With $ n< 0 $, it seems that we can consider only the inverse monomial inflation, a phenomenological model, which is discussed in \cite{Martin2, Ratra, Peebles, Huey} with a scalar potential as
\begin{eqnarray}
V(\varphi)= M^{4} (\dfrac{\varphi}{M_{4}})^{n}\,.
\end{eqnarray}
With this potential the slow-roll parameters are
\begin{eqnarray}
\epsilon_{v}\approx  \dfrac{\lambda M_{4}^{2}}{4\pi} \dfrac{V^{'2}}{V^{3}} \approx\dfrac{\lambda M_{4}^{2}}{4\pi}(\dfrac{\varphi}{M_{4}})^{n-2}\,,
\end{eqnarray}
\begin{eqnarray}
\eta_{v} \approx  \dfrac{\lambda M_{4}^{2}}{4\pi} \dfrac{V^{''}}{V^{2}} \approx\dfrac{n(n+1)}{4\pi M^{4}}\times (\dfrac{\varphi}{M_{4}})^{n-2}\,,
\end{eqnarray}
with the third slow-roll parameter $\xi_{v}^{2}$, which plays an important role in finding out the running of the spectral index, which is
\begin{eqnarray}
\xi_{v}^{2} \approx  \dfrac{\lambda^{2} M_{4}^{2}}{16\pi^{2}} \dfrac{V^{'}V^{'''}}{V^{4}} \approx\dfrac{\lambda^{2} n^{2}(n+1)(n+2)}{16\pi^{2} M_{4}^{2}}\times (\dfrac{\varphi}{M_{4}})^{2n-2}
\end{eqnarray}
 and $M_4$ is
\begin{eqnarray}
M_{4}\approx (N+\dfrac{n}{n-2}) \dfrac{\lambda n (n-2)}{4\pi} \bigg[6.0328\big(\dfrac{\lambda n^{2}(n-2)^{4}}{M_{4}^{4}}\big)^{1/6}\big( 58+\dfrac{n}{n-2}\big)^{2/3}\bigg]^{(n-2)}
\end{eqnarray}
where $ N $ is the number of e-folding. We get negative values for $ \epsilon_{v} $ and $ r $ for $ n=-1 $ and the relations blow up at $ n=-2 $ , so the high energy approximations of brane world gravity do not hold up this potential. Therefore, for inverse monomial inflation, the values of $ n $ must be less than $ -2 $. But the Brane-world corrections give a negligible tensor-to-scalar ratio and values of the spectral index $ n_{s}$ is pushed toward 1 ( scale-invariant spectrum). These results, however, are not consistent with the experimental results \cite{Planck2015, Hinshaw}.\\
Based on string theory, the typical inflation scenario from brane inflation can be realized via two effective mechanisms\cite{Dvali, Henry}: the slow-roll inflation and the Dirac-Born-Infeld (DBI) inflation\cite{DBI}. For the slow-roll mechanism, we consider the prototype and the KKLMMT model. A toy model of brane inflation, i.e., the prototype model, is a scenario that a pair of $Dp$ and $ \bar{D}p$-branes $ (p \geq 3)$ are put into the four large dimensions that are separated from each other in the extra six compactified dimensions. The inflation potential for this model is given by
\cite{Huang, Ma1, Ma2} 
\begin{eqnarray}
V=V_{0}\big(1-\dfrac{\mu^{n}}{\varphi^{n}})
\end{eqnarray}
where $V_{0}$ is an effective cosmological constant on the brane and the second term is an attractive force between the brane and anti-brane. The predictions of this model for $n=-2$ and $n=-4$ are still consistent with $ Planck+BK+BAO+H_{0}$ data\cite{Rui}.\\
A realistic brane inflation model is the KKLMMT model derived from the type $ IIB$ string theory \cite{Ma1, Ma2}. The inflation potential of this model is given by 
\begin{eqnarray}
V=\dfrac{1}{2}\gamma H^{2}\varphi^{2} + \dfrac{64 \pi^{2}\mu^{4}}{27}\big(1-\dfrac{\mu^{4}}{\varphi^{4}}\big)
\end{eqnarray}
where $H$ is the Hubble parameter and $ \gamma$ is the coupling between inflation $ \varphi$ and space expansion. Only the case of $ \gamma= 10^{-2}$ is marginally favored by considering the $ H_{0}$ measurement. The generalization of the KKLMMT model is called KKLTI with the potential of inverse harmonic function
\begin{eqnarray}
V_{KKLTI}=V_{0}\big(1+\dfrac{\mu^{n}}{\phi^{n}})^{-1}
\end{eqnarray} 
where at $ \mu \lesssim 1 $, a very good fit to the Planck 2018 data \cite{Planck2018} for $n=-4$ and an acceptable fit for $n=-2$ \cite{Linde2} are provided in the theory of $\alpha-$attractors \cite{Galante, Kallosh}. 
Therefore, it seems that the obtained range on $n<0$ in Eq.(\ref{sa}) is consistent with brane-world gravity based on the string theory due to corresponding measured strain sensitivity of the Adv.ligo and Virgo detectors during the time analyzed to determine the significance of GWGs. Also we can see  that the obtained most sensitive upper limits of frequency band
$\sim (20-100)$ Hz in recent paper \cite{pm} is within the frequency band $\sim (10^{-1}-10^{4})$ Hz of the correspond obtained ranges of $\beta$ in our work.  Hence, it is concluded that our result tells us that the string theory with its special models (prototype and KKLTI) may be a good candidates of potential  based on GWGs.     From other hand as mentioned in introduction, the analysis of NANOGrav \cite{qwq} finds strong evidence of a stochastic GWs, modeled as a power-law, with common amplitude and spectral slope
across pulsars. As one of the main source of stochastic GWs in the band 1-100 nHz is inflation \cite{qwq}. Therefore  these results may emphasize the evidence of
stochastic GWs that originated from inflation models such as prototype and KKLTI.

\section{discussion and conclusion}
 From the mentioned comparison, we explore the type of potential of inflation $\sim\phi^{n}$ for negative and positive $n$.  Based on measured strain sensitivity of GWGs and Planck data, our obtained constraints on $n$ and $\beta$ show that the negative $n$ is consistent than the positive one. Also, these new constraints correspond to prototype and KKTLI models, which are originated from string theory. Therefore the special models of string theory may be  a good candidates of potential. From other hand based on analysis of NANOGrav, these results may emphasize the evidence of
stochastic GWs that originated from inflation models such as  prototype and KKLTI.
Hence, gravitational waves play an important role in selecting the inflationary model and then data such as $ Planck+BK+BAO+H_{0}$ and NANOGrav will fine-tune the parameters of the model.
\appendix \label{m}
 \section {}
Under the slow-roll approximation, at the pivot wave number $ k_{0} $ the spectral parameters are given by \cite{Liddle, Liddle1, Liddle2}.
\begin{eqnarray}\label{n}
n_{t}\simeq &&-2\epsilon \nonumber\\
n_{s}\simeq &&1-6\epsilon +2\eta
\end{eqnarray}
In general, the spectral indices $ n_{t} $ and $ n_{s} $ are $ k$-dependent, described by the running parameters $ \alpha_{t} \equiv dn_{t}/d \ln k $ and $ \alpha_{s} \equiv dn_{s}/d \ln k $, respectively \cite{Miao,Liddle, Liddle1, Liddle2, Koso, Solokhin }.
The nonzero $ \alpha_{s} $ would induce an $ n_{s} $ greater than one. The value of $ n_{t} $ is quite uncertain, but $ n_{s} $ can be well constrained by seven-year WMAP, $ n_{s}=0.967\pm 0.014 $ and $ n_{s}=0.982^{+0.020}_{-0.019} $ \cite{Komatsu}. Independently, SPSS III shows $ n_{s}=0.96\pm 0.009 $ \cite{Sanchez} and the Planck 2018 reports the scalar spectral index from $ n_{s}=0.9626\pm 0.0057 $ to $ n_{s}=0.98\pm 0.015 $ \cite{Planck2018}.\\ 
The ratio of the primordial tensor power spectrum to the scalar power spectrum is defined based on the Eq. (\ref{ratio}) as
\begin{equation}
r\equiv\dfrac{\Delta_{h}^{2}(k, \eta_{\star})}{\Delta_{R}^{2}(k, \eta_{\star})}=16\epsilon
\end{equation}
At the pivot number $ k_{0} $, it will be 
\begin{equation}\label{13}
r \simeq \dfrac{\Delta_{h}^{2}(k_{0})}{\Delta_{R}^{2}(k_{0})}
\end{equation}
With above approximation (e.g., (\ref{13})) and CMB observation, one find a simple relation
\begin{equation}\label{beta}
n_{t}=2\beta +4\,,
\end{equation}
 According to Eqs. (\ref{epsilon}) and (\ref{ratio}), with straightforward calculations one can obtain \cite{to}
\begin{equation}\label{r}
r=\dfrac{8n}{n+2}(1-n_{s})
\end{equation}
Then, based on eqs.(\ref{n}) to (\ref{beta}) one has
\begin{equation}
\beta=-2-\dfrac{n}{2(n+2)}(1-n_{s})
\end{equation}

\end{document}